\documentclass[a4paper,fleqn,usenatbib]{mnras}

\usepackage{newtxtext,newtxmath}

\usepackage[T1]{fontenc}
\usepackage{ae,aecompl}

\usepackage{graphicx}  
\usepackage{amsmath}   
\usepackage{subfig}

\newcommand{\ccm}{\,cm$^{-3}$}
\newcommand{\gccm}{\,g cm$^{-3}$}
\newcommand{\kms}{\,km s$^{-1}$}
\newcommand{\fcyl}{f_\text{cyl}}

\title[Taking off the edge]{Taking off the edge - simultaneous filament and end core formation}

\author[S. Heigl et al.]{
S. Heigl,$^{1,2}$\thanks{E-mail: heigl@usm.lmu.de}
E. Hoemann,$^{1,3}$
A. Burkert,$^{1,2,3}$
\\
$^{1}$Universit\"ats-Sternwarte, Ludwig-Maximilians-Universit\"at M\"unchen, Scheinerstr. 1, 81679 Munich, Germany\\
$^{2}$Excellence Cluster ORIGINS, Boltzmannstrasse 2, 85748 Garching, Germany\\
$^{3}$Max-Planck Institute for Extraterrestrial Physics, Giessenbachstr. 1, 85748 Garching, Germany\\
}

\date{Accepted XXX. Received YYY; in original form ZZZ}

\pubyear{2021}

\begin{document}
\label{firstpage}
\pagerange{\pageref{firstpage}--\pageref{lastpage}}
\maketitle

\begin{abstract}
  Simulations of idealised star-forming filaments of finite length typically
  show core growth which is dominated by two cores forming at its respective
  end. The end cores form due to a strong increasing acceleration at the
  filament ends which leads to a sweep-up of material during the filament
  collapse along its axis. As this growth mode is typically faster than any
  other core formation mode in a filament, the end cores usually dominate
  in mass and density compared to other cores forming inside a filament.
  However, observations of star-forming filaments do not show this
  prevalence of cores at the filament ends. We explore a possible
  mechanism to slow the growth of the end cores using numerical
  simulations of simultaneous filament and embedded core formation, in our
  case a radially accreting filament forming in a finite converging flow.
  While such a setup still leads to end cores, they soon begin to move inwards
  and a density gradient is formed outside of the cores by the continued
  accumulation of material. As a result, the outermost cores
  are not longer located at the exact ends of the filament and the density
  gradient softens the inward gravitational acceleration of the cores.
  Therefore, the two end cores do not grow as fast as expected and thus do not
  dominate over other core formation modes in the filament.
\end{abstract}




\section{Introduction}
\label{sec:introduction}

  A vital step in the star-formation process is the fragmentation of gas
  from molecular clouds on scales of a few tens of parsecs down to
  individual star-forming cores with sizes of a few tenths of a parsec.
  \citet{schneider1979} and \citet{larson1985} established the idea that
  this collapse does not happen directly but goes through an intermediate
  phase of filamentary fragmentation. Indeed, the dust observations of the
  Herschel space telescope \citep{andre2010} showed that molecular clouds
  are dominated by intricate networks of filaments which also are the
  locations where most cores are found \citep{arzoumanian2011, konyves2015}.

  Many processes related to filament formation and their subsequent
  fragmentation are still unclear. Filamentary structure has been seen as
  a natural outcome of turbulent box simulations, with or without gravity
  \citep{vazquez1994, klessen2000, padoan2001, moeckel2015, federrath2016}.
  These simulations have shown that they form by the compression of gas in
  the crossing of two planar shocks or as a consequence of dissipation at
  the end of the turbulent cascade \citep{kritsuk2013, smith2014, smith2016}.
  The formation sites of stars are commonly situated in the densest
  regions where several filaments overlap with only few cores forming at the
  end of filamentary structure \citep{girichidis2014, smith2014, federrath2016}.
  A similar morphology can be observed in collapsing cloud simulations
  \citep{dale2011, bate2012, gomez2014} where the global contraction usually
  leads to a central dense filament regulating the flow of gas from the
  cloud to the cores.

  Nevertheless, simulations of individual, isolated filaments are
  dominated by a different morphology
  \citep{bastien1983, bastien1991, clarke2015, seifried2015} with strong
  overdensities forming at both ends of the filament. This is due to the
  gravitational acceleration being the largest at the filament ends
  \citep{burkert2004, pon2011, pon2012, toala2012, clarke2015}, a
  concept which can be more generalised to more complex structures and
  which is known as so-called "gravitational focusing" or "edge-effect"
  \citep{burkert2004, hartmann2007, li2016},
  where the magnitude of the gravitational acceleration scales with the
  curvature of the surface.

  While there are observed cases of end dominated filaments
  \citep{zernickel2013, kainulainen2016, johnstone2017, ohashi2018,
  dewangan2019, bhadari2020, yuan2020, cheng2021}, the majority of observed
  filaments do not show a particular disposition to forming their most
  massive cores at their ends which can be due to several reasons.
  For instance, filaments forming in networks are interconnected and
  therefore do not have the density gradients at their ends required to
  form end cores. Nevertheless, one would expect the ends of the network
  to enhance core formation. Moreover, it has been shown that if filaments
  form with low aspect rations, large initial central overdensities or via
  filament mergers, the collapse of the gas is concentrated onto its centre
  \citep{keto2014, seifried2015, hoemann2021}, a morphology which has also
  been observed in large line-mass filaments \citep{kirk2013, henshaw2014}.
  Large central overdensities are however not expected in low-mass star
  forming filaments.

  The discrepancy between theory and observations leads to the
  question whether different flow patterns in the formation of filaments can
  suppress dominant cores at the filament ends. In this paper we present
  such a process: the simultaneous migration of the forming end cores
  while the filament is under formation itself. In the
  turbulent formation regime, filaments form by colliding shocks leading to
  mass inflow on the cross-section of the shocks over a timescale which is
  given by the size of the inflow region and the shock velocity.
  While planar shocks first lead to the formation of sheets, these can
  further collapse and form filaments with sustained accretion along the plane
  of the sheet \citep{shimajiri2019, chen2020}. Moreover, direct gravitational
  collapse of cylindrically distributed material onto its centre will lead to
  a sustained radial inflow of gas \citep{clarke2017, heigl2020} which can
  also be driven by the gravitational potential of the filament itself. While
  a purely gravitationally driven inflow makes the accretion of material outside
  of the end cores more unlikely, the process of simultaneous mass inflow and
  core migration can still occur. In our idealised simulations we assume a
  continuous radial inflow which replenishes the material outside of the end
  and stabilises the filamentary structure.

  Accretion flows have been observed in many filaments
  \citep{schneider2010, kirk2013, palmeirim2013, shimajiri2019, bonne2020}
  and typically show estimated accretion rates of around
  $10-100$ M$_\odot$ pc$^{-1}$ Myr$^{-1}$ and accretion velocities of the
  order of $0.25-1.0$ \kms. If we assume a shock inflow region feeding the
  filament of around 0.5 pc, this would lead to a sustained inflow of
  material on timescales of 0.5-2 Myr. We demonstrate below that, while a
  filament is forming in such a sustained accretion flow, end cores will
  indeed begin to form and start migrating inwards while new material
  reforms the filament behind them. As we will show, this results in a
  suppression of the runaway growth of these cores and to core formation
  within the whole filament on roughly the same timescale.

  The sections of the paper are organised as follows: \autoref{sec:concepts}
  discusses the basic physical principles which lead to the formation of
  cores at the filament ends. The numerical principles and initial conditions
  of the simulations are introduced in \autoref{sec:numerics} and their
  results are presented and discussed in \autoref{sec:simulations} and
  \autoref{sec:cores}. We draw the conclusions and summarise the results in
  \autoref{sec:discussion}.

\section{Basic concepts}
\label{sec:concepts}

  While the edge-effect is not the only mode of core formation in filaments,
  it usually is the fastest. The end cores form at the position of the
  strong density gradient located at the ends of a filament due to the
  sharp increase in gravitational acceleration.
  Assuming a filament has a
  constant density $\rho$, a total length of $L$, a radius $R$ and that
  the filament is directed along the x-axis, this acceleration is given by
  \citep{burkert2004}:
  \begin{equation}
    a_x = -2\uppi G\rho\left[2x - \sqrt{R^2+\left(L/2 + x\right)^2}+
          \sqrt{R^2+\left(L/2 - x\right)^2}\right]
    \label{eq:acc}
  \end{equation}
  with $G$ being the gravitational constant. The acceleration on the ends
  increases steeply for large aspect ratios which can be seen, for instance,
  in \citep{pon2012}. For $x=L/2$ and large lengths $(L \gg R)$ this
  acceleration equals:
  \begin{equation}
    a_x = -2\uppi G\rho R.
  \end{equation}
  In fact the acceleration increase is even stronger for more centralised
  radial profiles \citep{hoemann2022} such as the isothermal hydrostatic
  cylindrical profile \citep{stodolkiewicz1963, ostriker1964}:
  \begin{equation}
    \rho(r) = \frac{\rho_\text{c}}{\left(1+\left(r/H\right)^{2}\right)^{2}}
    \label{eq:ost}
  \end{equation}
  where $r$ is the cylindrical radius and $\rho_\text{c}$ is its central
  density. The radial scale height $H$ is given by the term:
  \begin{equation}
    H^2 = \frac{2c_\text{s}^2}{\uppi G \rho_\text{c}}
  \end{equation}
  where $c_\text{s}$ is the isothermal sound speed. This profile can only
  support a maximum mass per length in hydrostatic equilibrium. The maximum
  value is calculated by integrating the profile radially to infinity:
  \begin{equation}
    \left(\frac{M}{L}\right)_\text{crit} = \frac{2c_\text{s}^2}{G}
    \label{eq:lmcrit}
  \end{equation}
  and one can define the parameter $\fcyl$ which gives the value of the
  current line-mass compared to the critical value \citep{fischera2012}:
  \begin{equation}
    \fcyl = \left(\frac{M}{L}\right)/\left(\frac{M}{L}\right)_\text{crit}.
  \end{equation}
  In pressure equilibrium with an ambient pressure, the hydrostatic radius
  of the isothermal filament is given by
  \begin{equation}
    R = H \left(\frac{\fcyl}{1-\fcyl}\right)^{1/2} = \left(\frac{2c_\text{s}^4}{\uppi G p_\text{ext}}\left(\fcyl(1-\fcyl)\right)\right)^{1/2}
    \label{eq:rad}
  \end{equation}
  as the central and outer density are connected by
  \begin{equation}
    \rho_c = \frac{\rho(R)}{(1-\fcyl)^2}.
    \label{eq:rhoc}
  \end{equation}

\section{Numerical set-up}
\label{sec:numerics}

  \begin{figure}
    \centering
    \includegraphics[width=0.8\columnwidth]{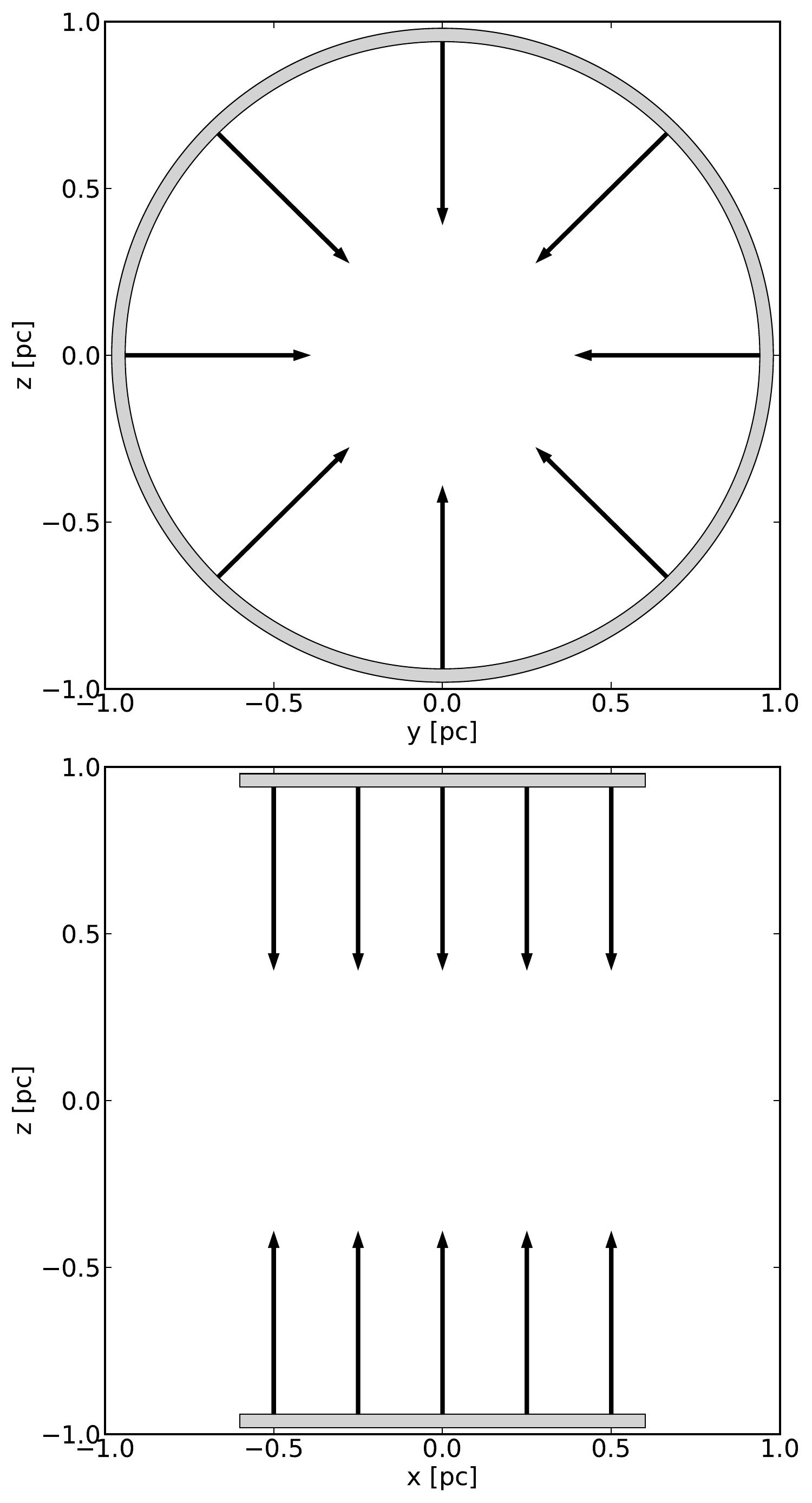}
    \caption{Schematic diagram of the inflow region as defined
    in the initial conditions. We show central cuts through the simulation
    box in the y-z and x-z plane. The inflow region where material is
    constantly replenished is given by the gray shaded region and the inflow
    velocity given to the gas is shown as arrows.}
    \label{fig:sketch}
  \end{figure}

  \begin{figure*}
    \includegraphics[width=2.0\columnwidth]{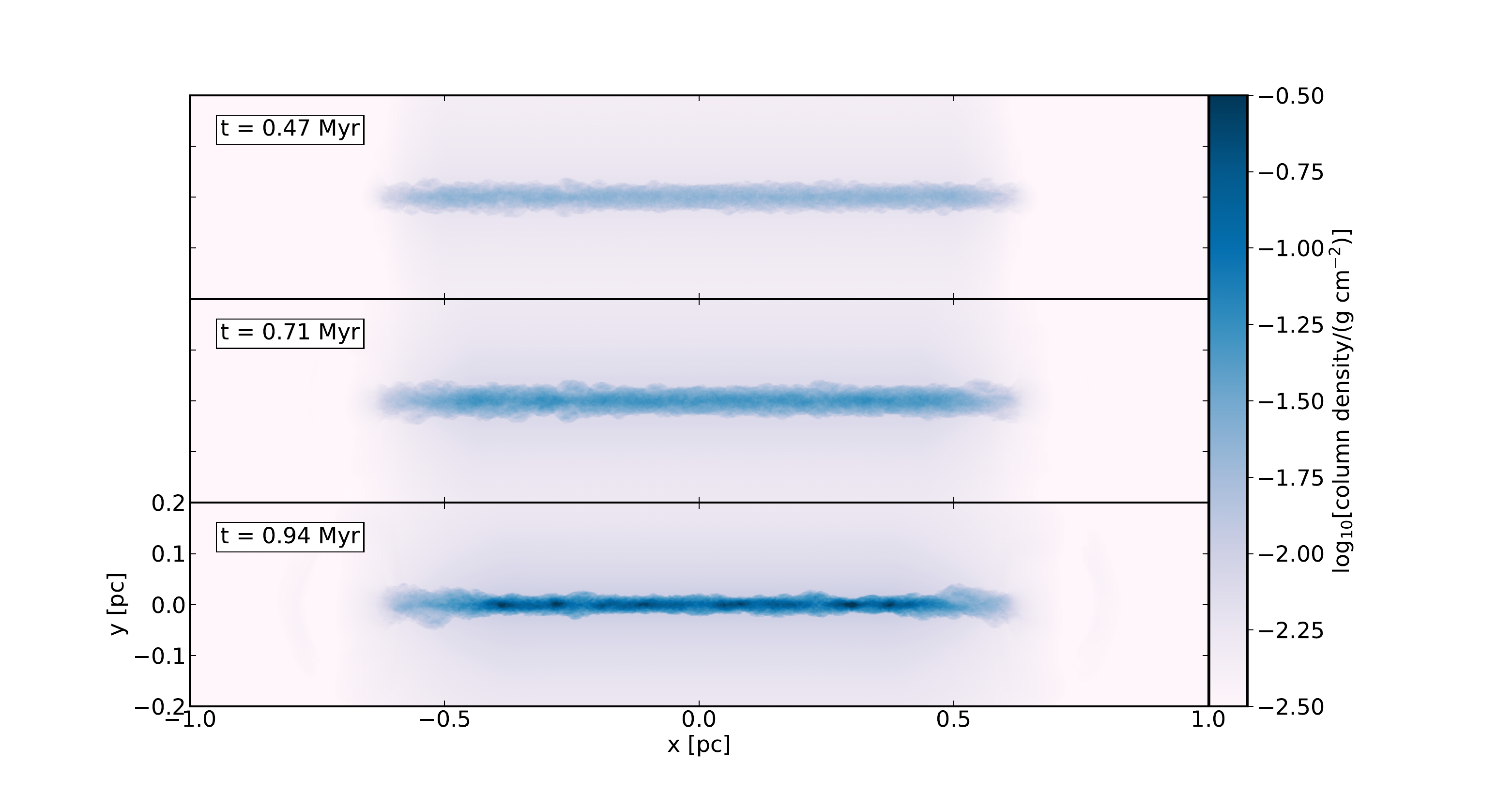}
    \caption{Overview of the simulation given by the density projection
    along the z-axis at different points in time. The radial accretion flow
    forms a turbulent filament which forms regularly spaced cores along its
    axis. The y-axis label is the same for every panel.}
    \label{fig:pro}
  \end{figure*}

  We ran our simulation with the code \textsc{ramses}
  \citep{teyssier2002} using a second-order Godunov scheme
  to solve the conservative form of the discretised Euler equations
  on an Cartesian grid. We applied the MUSCL scheme \citep[Monotonic
  Upstream-Centred Scheme for Conservation Laws][]{vanLeer1979}
  in combination with the HLLC-Solver
  \citep[Harten-Lax-van Leer-Contact][]{toro1994}
   and the multidimensional MC slope limiter
  \citep[monotonized central-difference][]{vanLeer1977}.

  The formation and gravitational evolution of a finite-sized filament is
  simulated taking into account mass accretion from the surrounding. This
  is a key difference to previous simulations \citep{bastien1983,
  bastien1991, clarke2015, seifried2015} that assumed an already
  existing filament and that followed its subsequent collapse.
  We do this by setting up a cylindrical inflow zone with a
  radius that is set in the y-z dimension and define a constant radial
  accretion flow onto the central x-axis, similar to the initial conditions
  of \citet{heigl2018} and \citet{heigl2020}. In contrast to the former
  cases however, all boundaries of the box are set to open and we do not use a
  periodic boundary condition in x direction. In addition we limit the
  accretion flow to the central section of the box with a size of 0.6 times
  the box length. We choose the box to have a physical size of 2.0 pc
  resulting in an accretion zone with a radius of 1.0 pc and a length of
  1.2 pc. A schematic diagram of inflow can be seen in \autoref{fig:sketch}
  where we show central cuts through the simulation box of the y-z and x-z
  plane.

  The gas is set to be isothermal with a temperature of $10$ K
  with a molecular weight of $\mu = 2.36$. In order to break the
  symmetry of the simulation, we introduce random density perturbations
  in the initial density and in the inflow following a flat distribution with
  a maximum value of 10\% which allows the driving of turbulent motions.
  As shown in \citet{heigl2018}, the amplitude of the turbulence does not
  depend on the strength of the initial perturbation even when varied
  over close to four orders of magnitude ranging from 0.01\% up to 50\%. We do
  not set any preferred initial perturbation scale but vary every cell and let
  the cores form self-consistently from the resulting turbulence.

  We set the mass accretion to a constant rate which is defined
  by the distance of the inflow zone to the central axis $R_\text{0}$,
  its density $\rho_\text{0}$ and the accretion velocity $v_a$:
  \begin{equation}
     \frac{\dot{M}}{L} = 2\pi \rho_0 R_0 v_a,
  \end{equation}
  where both of the latter values are renewed at every timestep in
  the simulation. The constant mass accretion set at the radius $R_\text{0}$
  propagates inward and leads to a radial density profile outside of the
  forming filament of:
  \begin{equation}
    \rho(r) = \frac{\rho_\text{0}R_\text{0}}{r}.
    \label{eq:acc}
  \end{equation}
  We vary the value of the accretion rate in different simulations
  by adjusting the density $\rho_\text{0}$ and the accretion velocity $v_a$.

  As the filament diameter itself is only a fraction of the total box size
  and as we are not interested in the details of the accretion flow itself,
  we employ adaptive mesh refinement (AMR) in order to speed up the simulation
  by increasing the resolution inside the filament and keeping the resolution
  low in the ambient medium. We vary the resolution level from 8 to 12 which
  means that our base grid has a resolution of $256^3$, corresponding to a
  cell size of $7.8\times10^{-3}$ pc, whereas our filament is resolved by
  a resolution of $4096^3$, corresponding to a cell size of
  $4.9\times10^{-4}$ pc. As a refinement condition we use the Truelove
  criterion \citep{truelove1997} increasing the resolution as soon as the
  Jeans length is not resolved by 256 cells. We choose an aggressive
  refinement value of 256 in order to guarantee that the filament is
  resolved fully with our highest resolution. We stop the simulation as soon
  as a core collapses and we do not fulfill the Truelove criterion for our
  highest refinement level anymore.

  \begin{figure}
    \includegraphics[width=1.0\columnwidth]{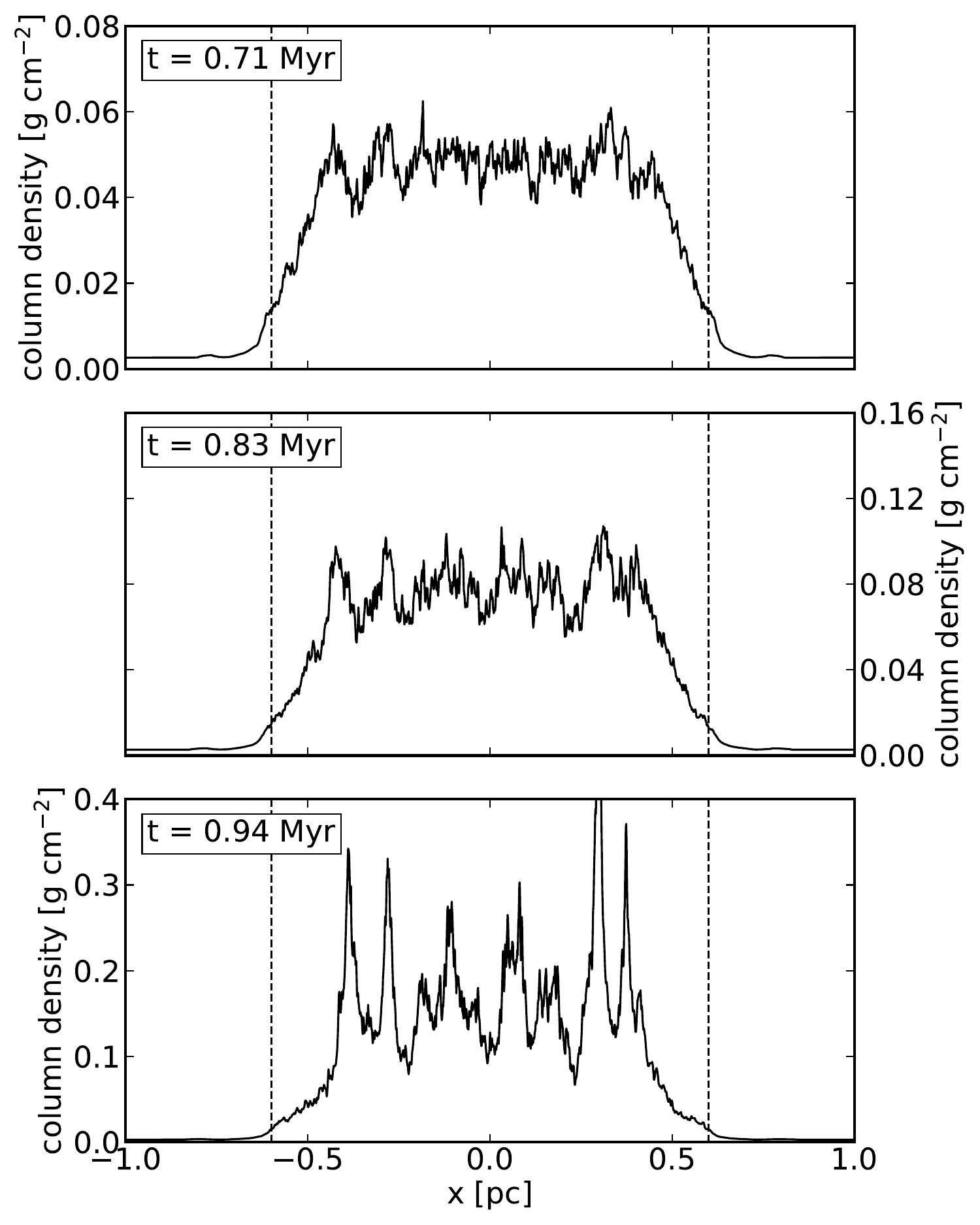}
    \caption{Time evolution of the column density in z direction along the
    central x-axis. The inflow region is marked by the vertical dashed lines.
    In contrast to developing dominant end cores, the inner dense region
    stays relatively even. The filament contracts over time and
    forms equally dense cores along its axis. The x-axis label applies to
    every panel.}
    \label{fig:pline}
  \end{figure}

\section{General results}
\label{sec:simulations}

  As a fiducial case study we set the initial density in the box and that
  of the accretion flow to a mean value of $3.92 \times 10^{-22}$ \gccm,
  corresponding to a number density of about $100$ \ccm. The inflow velocity is
  set to a value of $0.75$ \kms, which is equal to a Mach number of 4.0 for 10 K
  gas. The resulting accretion rate is therefore about $27.9$ M$_\odot$ pc$^{-1}$ Myr$^{-1}$
  which is consistent with predicted and observed accretion rates and velocities
  \citep{schneider2010, kirk2013, palmeirim2013, shimajiri2019, bonne2020}. However,
  as we do not include an initial accretion profile of the form of \autoref{eq:acc} in
  this case, the accretion rate will be lower in the beginning and will grow to the
  constant value as soon as the accretion profile is established.

  Here we provide a detailed analysis of our general findings
  starting with an overview of the evolution of the filament in \autoref{fig:pro}.
  We plot the column density along the z-axis assuming our gas is optically thin.
  The different panels show different time steps of the simulation. As the
  material streams to the box centre it forms a filament which gains mass
  over time. In addition, the inflow stabilises the filament
  against radial expansion. One can clearly see the turbulent nature of
  the gas which is driven by the dissipative accretion shock and a smooth
  density gradient at the two end points of the filament ends extending into
  the ambient medium. A tentative imprint of the inflow region can be seen in
  column density as slightly darker region around the filament itself. Other
  than its straight form due to the nature of how we defined the filament
  accretion, the visual impression closely resembles observed filamentary
  structure.

  While a pre-defined, non-accreting filament would immediately start forming
  end cores, most of the evolution of the accreting filament is uneventful. As
  it grows in mass over time it becomes denser while retaining a relatively
  soft gradient in column density to the ambient medium. After about 0.7 Myr,
  cores condense out of the filament material with none being particularly
  dominant compared to the others. The last snapshot of the simulation is
  shown in the third panel where one can see several cores
  along the filament with similar spacing. Note that the outermost
  cores do not coincide with the filament ends.

  In addition to the slices, we also show column density cuts along the
  central axis in \autoref{fig:pline} where the evolution of the
  overdensities can be seen in more detail. The first panel shows the
  earliest stage of overdensities forming in the filament at 0.71 Myrs.
  Although material is accreted in the region between $x=-0.6$ and $x=0.6$,
  one can already see that the densest part is shorter and forms a central
  dense region which contracts over time. We will concentrate on
  this region, which we will address as dense inner region, in further
  dynamical investigations as it contains most of the filament material.
  While for a non-accreting filament a contraction along the filament axis
  would lead to an enhancement of material in the filament ends, in this
  case the accreting filament shows a homogeneous increase along its axis
  with only small density fluctuations. The next panel at 0.83 Myrs
  shows a time step where several cores have formed. While there are also two
  overdensities forming at the end of the inner dense region,
  these do not form earlier than the rest and have similar column densities as
  the other cores. While the inner dense region contracts along the filament
  axis, one can see that the continuous inflow brings in new material around
  its ends. The last panel at at 0.94 Myrs shows the last time
  step of the simulation where one of the cores begins to collapse leading to
  a drastic increase in its central column density. Interestingly, it is not
  one of the outermost ones as one would expect for an end dominated
  filament but the second from the right. Again, neglecting the collapsing one, all
  cores have similar central column densities albeit with a possible tendency
  of the outer cores to be slightly denser.

   \begin{figure}
    \includegraphics[width=1.0\columnwidth]{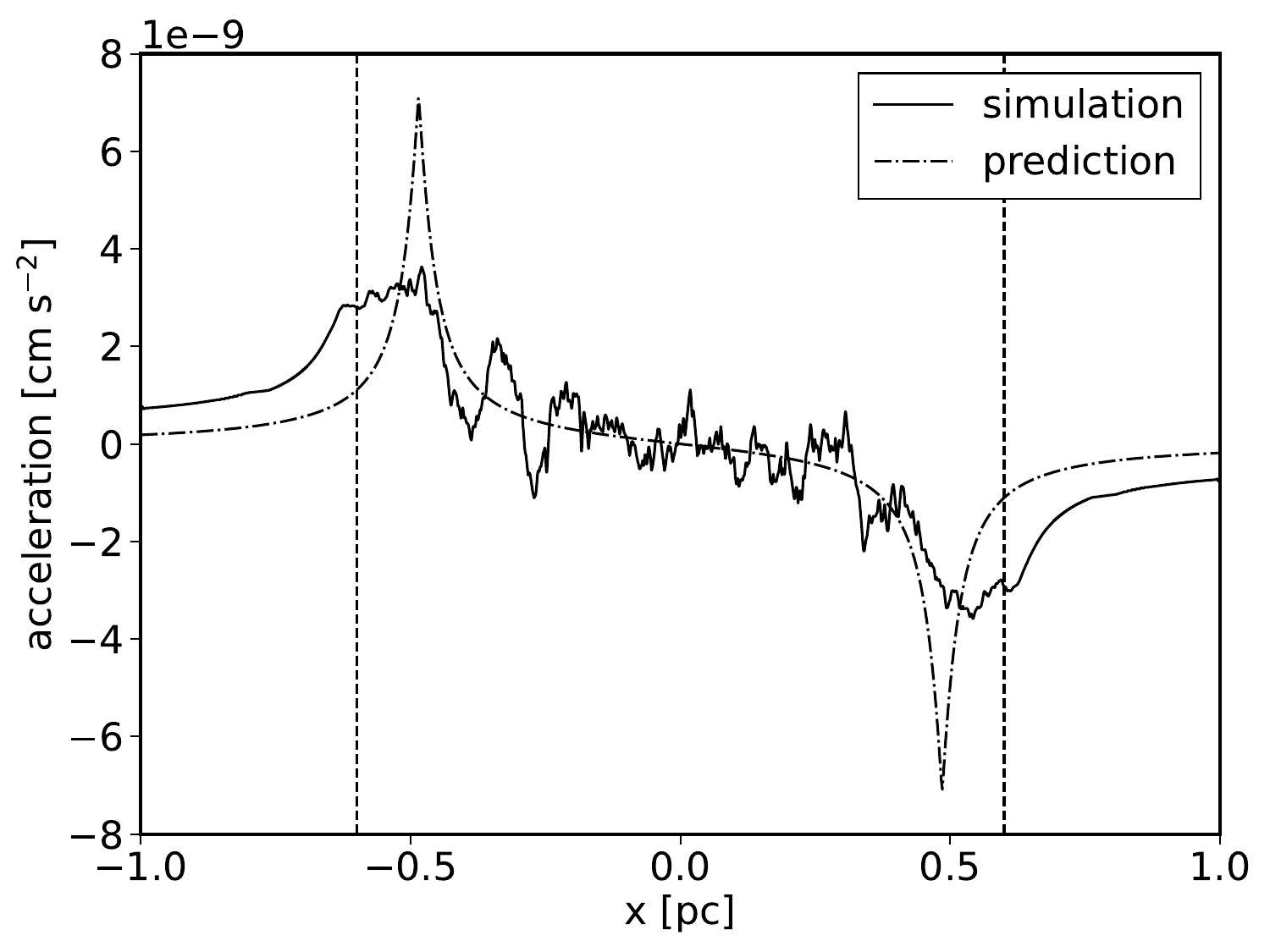}
    \caption{Gravitational acceleration along the filament axis for t=0.83
    Myr. The solid curve shows the measured acceleration in the simulation
    while the dashed-dotted curve is the analytic prediction of
    \autoref{eq:acc} for the dense inner region. The inflow region is marked
    by the dashed vertical lines. There is a considerable softening of the
    acceleration at the positions of the end cores.}
    \label{fig:acc}
  \end{figure}

  The reason for the difference in the importance of the end cores
  compared to the collapse of an non-accreting filament has to lie in the
  acceleration at the filament ends. We therefore analyse the output at
  0.83 Myrs in more detail. First, we plot the measured gravitational
  acceleration along the filament in the simulation as the solid curve
  in \autoref{fig:acc} together with
  the analytical prediction of \autoref{eq:acc} where we use the length,
  the mean density and the mean radius of the dense inner region. We
  define the length of the dense inner region by the distance between
  the outer edges of the outermost cores which we find by determining the two
  points where the column density first exceeds the mean column density
  of the dense inner region calculated in a section between -0.25 pc and 0.25 pc.
  As the column density is relatively flat, taking the mean of an inner
  section smoothes out possible over- and underdensities. Within the dense
  inner region the prediction agrees very
  well with the measured acceleration. However, at its ends, where the end
  cores form, it differs strongly. The material outside of the outermost cores
  leads to a softening of the gravitational attraction at the position of the
  end cores. As the ends of the inner region are more and more embedded in
  newly accreted material, the acceleration at their position matches more a
  region contained inside a filament. As the new density gradient is very
  shallow, the new ends are evenly accelerated instead of leading to a sweep-up
  of material.

\begin{figure}
    \includegraphics[width=1.0\columnwidth]{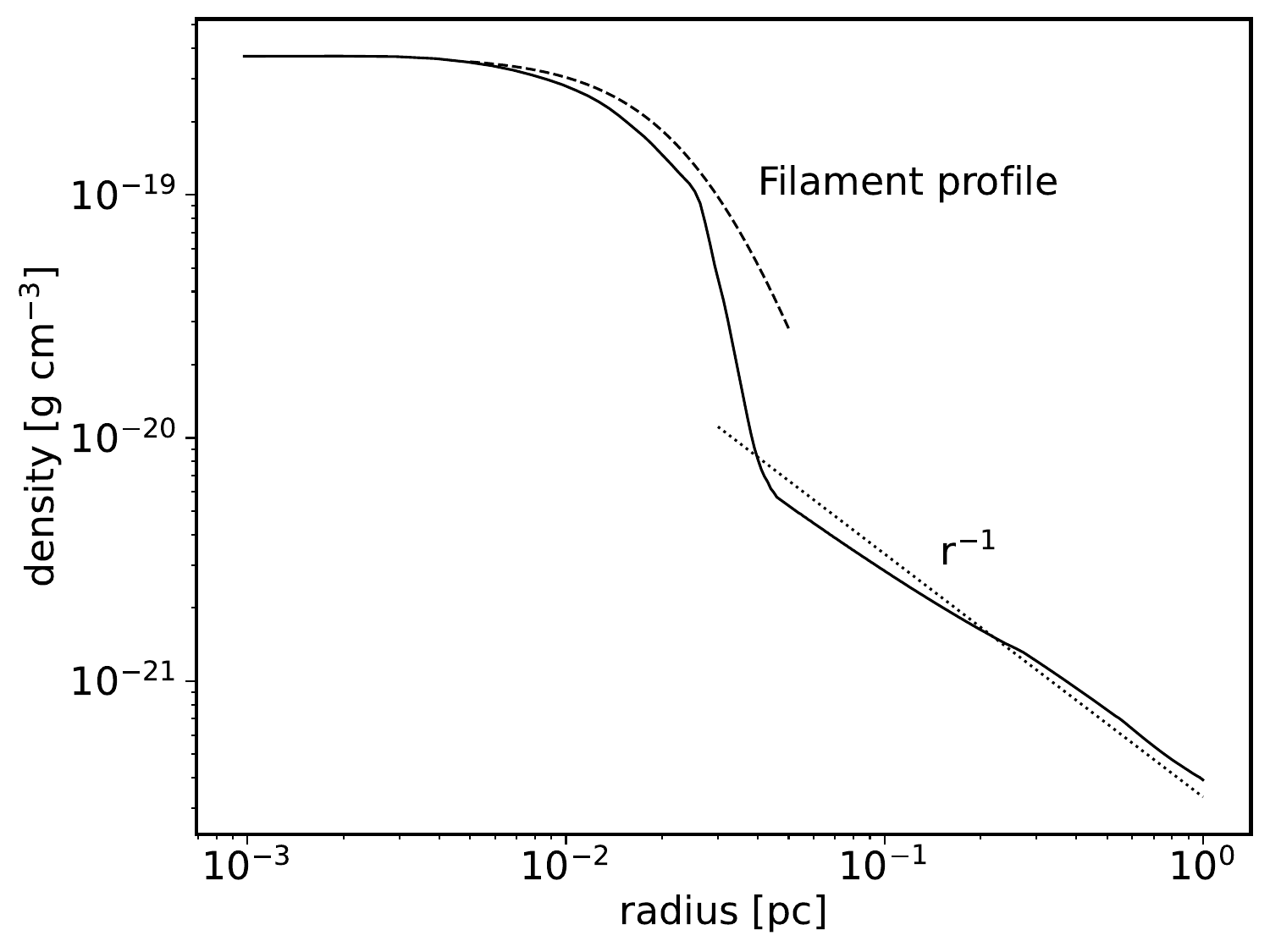}
    \caption{Radial density profile of the filament and the
    surrounding accretion region at 0.83 Myrs. We average the density along
    the filament axis inside the dense inner region. The filament follows
    the hydrostatic solution of \autoref{eq:ost} which is defined by its central
    density and shown by the dashed line. The accretion region follows an
    $r^{-1}$ profile in agreement for a constant accretion rate set by the boundary
    values in the accretion zone as given in \autoref{eq:acc}.}
    \label{fig:profile}
  \end{figure}

  In general, this result should be independent of the density profile
  of the filament as a soft filament end should always decrease the
  acceleration at the end of a filament. However, as we are also interested
  in the question whether we can apply analytical perturbation theory which
  has been established for the hydrostatic profile of \autoref{eq:ost}, we
  furthermore plot the density profile perpendicular to the filament radially
  averaged over the dense inner region in \autoref{fig:profile}. In the plot one
  can discern two distinct regions which are composed of the filament itself
  and the accretion region. The accretion region follows an $r^{-1}$ profile as
  established in \autoref{eq:acc} until it hits the filament surface where
  an accretion shock is formed and a jump in density occurs. The filament
  itself follows well the hydrostatic solution which is defined by the
  central density and given in the plot by the dashed black line. It does
  so despite the filament containing turbulent motions as was already the
  case in \citet{heigl2020} as the turbulence only adds a constant background
  pressure. We will use this fact for analysing core distances in the next
  section.

  \begin{figure*}
    \includegraphics[width=2.0\columnwidth]{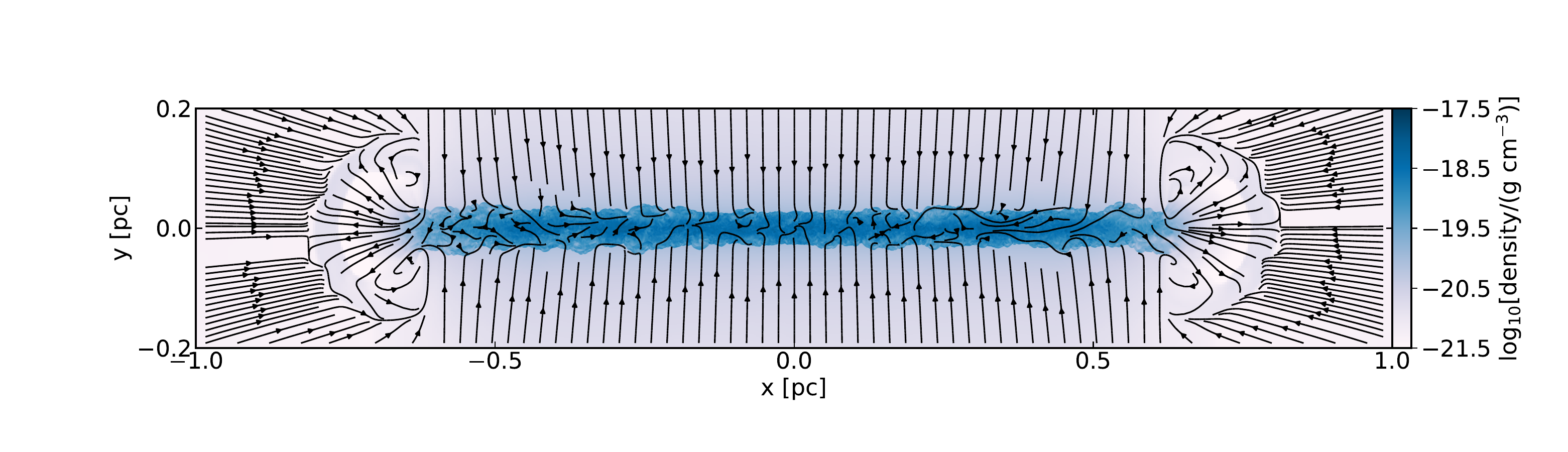}
    \caption{Density slice through the centre of the filament at 0.83 Myr.
    The streamlines show the direction of the velocity in the plain of the
    slice. One can see that material at the filament ends is flowing into the
    ambient medium where it forms shock fronts.}
    \label{fig:slice}
  \end{figure*}

  An interesting consequence of the radial compression of the gas is that a
  small amount of material is pushed out at the filament ends due to the
  pressure gradient from the dense compressed material to the ambient medium.
  This leads to a small shocked region surrounding both filament ends.
  Although this feature is very hard to detect in the column density of
  \autoref{fig:pro}, it can be seen in the density slice through the centre
  of the filament shown in \autoref{fig:slice}. In addition to the density,
  we also show the direction of the velocity vectors in the plane of the
  slice overlaid as streamlines. One can see the inflow region accreting onto
  the filament, the ambient medium being gravitationally pulled towards the
  filament and the two outflow regions pushing against it. While the density
  of the outflow regions is very thin, shock fronts form where they hit
  and compress the ambient medium. Wether such outflow features can be
  detected in observations is an interesting question, its thin density
  however makes such observations challenging. Note that the outflows do not
  affect the overall evolution of the filament itself as the mass loss rate
  of about $\sim 0.1$ M$_\odot$ Myr$^{-1}$ at both ends is negligible small
  compared to the total mass accretion rate of $27.9$ M$_\odot$ Myr$^{-1}$.

\section{Core separation}
\label{sec:cores}

  A key question of filamentary fragmentation theory and of star formation is
  the existence of a quasi-regular core spacing. This mode of fragmentation
  has been predicted by hydrostatic models, as there exists a dominant mode
  for which the growth time scale of the overdensities is the shortest
  \citep{stodolkiewicz1963, larson1985, nagasawa1987, inutsuka1992,
  nakamura1993, gehman1996}. Observations of core spacing have been
  inconclusive with several studies showing lower separations than expected
  \citep{andre2010, kainulainen2013, teixeira2016} with some agreeing with
  the predictions \citep{beuther2015, contreras2016, kainulainen2016}.
  Consequently, the core separation in our simulations is of considerable
  interest.

  The predicted dominant mode of fragmentation in filaments
  is calculated from the polynomial approximation
  \begin{equation}
    y(\fcyl) = \sum_{i=0}^5 a_i \fcyl^{i/2}
    \label{eq:dom}
  \end{equation}
  with the respective values $a_i$ shown in table E.1 of \citet{fischera2012}
  which are also listed in \autoref{table:constants}. Here, y represents either
  the length scale $\lambda_{dom}$, the dominant fragmentation length in units
  of 1/FWHM, with FWHM being the full width half maximum of the filament profile,
  or the length scale FWHM itself in units of 1/$H$. This means we can express
  the dominant fragmentation length as function of the current average line-mass
  $\fcyl$  and the current average scale height $H$ and therefore by extension the
  current average central density. This leads to a major complication for verifying the
  theoretical predictions, not only in simulations, but also in observations. In
  an accreting filament the average line-mass constantly increases and the average
  central density changes over time. However, one can only measure the current average
  line-mass and the current average central density at a single point in time and
  not the line-mass and central density that the filament had when the cores began forming.
  While in principle one could follow the core growth in simulations from the beginning,
  it is impossible to measure the core separation at their time of birth directly
  as the cores are indiscernible from other turbulent perturbations at this point
  of time. Moreover, a further complication is that the length of the filament and
  consequently the distance between cores will decrease over time due to the
  gravitational collapse along its axis. At a single point in time, one can only
  measure the current spacing.

  However, if the cores in the simulation all began to form at the same time and
  thus all formed at the same dominant fragmentation length, we can test if we can
  determine this point in time. While the cores in principle can form independently
  at any time there are two key observations which lead to the assumption that
  all cores are seeded on the same dominant wavelength:
  \begin{itemize}
    \item the cores seem all to appear at the same time and evolve on similar
      timescales
    \item we generally do not see any cores sub-fragmenting and merging
  \end{itemize}
  This means that the number of cores which are present at the end of the simulation
  are a consequence of the imprint of the singular dominant wavelength when the
  cores began to grow. Even as their distance between each other lessens over
  time due to the collapse of the dense inner region, we can backtrace their
  mean separation, determined by the number of cores and the length of the
  dense inner region, in order to see if it matches the prediction of the
  co-evolving dominant fragmentation length.

  \begin{table}
    \centering
    \caption{Constants of the polynomial approximation of the predicted dominant fragmentation length for
    the values calculated in \citet{nagasawa1987} as listed in \citet{fischera2012}. FWHM is the full
    width half maximum of the filament profile and $\tau_\text{dom}$ is the growth time of the
    dominant mode in units of $\sqrt{4\pi G \rho_c}$.}
    \label{table:constants}
    \resizebox{\hsize}{!}{
    \begin{tabular}{c c c c c c c}
      \hline\hline
      & $a_0$ & $a_1$ & $a_2$ & $a_3$ & $a_4$ & $a_5$ \\
      \hline
      $\lambda_\text{dom}/\text{FWHM}$ & 6.25 & 0.00 & -6.89 & 9.18 & -3.44 & 0.00 \\
      $\text{FWHM}/H$ & 0.00 & 1.732 & 0.00 & -0.041 & 0.818 & -0.976 \\
      $\tau_\text{dom}\sqrt{4\pi G \rho_c}$ & 4.08 & 0.00 & -2.99 & 1.46 & 0.40 & 0.00 \\
      \hline
    \end{tabular}
    }
  \end{table}

  With the purpose of conducting a statistical study on the number of cores which
  form in accreting filaments, we perform a new set of simulations. In contrast to
  the fiducial case, we now introduce several changes in order to keep the mass accretion
  rate as close to constant as possible which enables us to map the line-mass to a
  certain time in the simulation more easily. Therefore, the velocity and density
  of the material reaching the filament have to be kept constant. For low accretion
  Mach numbers, the gravitational acceleration acting upon the gas starting at the
  edge of the box with a size 2.0 pc can already increase the accretion velocity
  significantly once the gas reaches the filament. We prevent this by using a larger
  accretion Mach number of 6.0 for which we do not see any significant increase in velocity.
  Furthermore, as the density and the mass accretion rate of the ambient medium
  is only constant if the density profile follows \autoref{eq:acc}, we also include
  it as the initial density profile at the start of the simulation. In order to
  obtain a better statistical sample of the number of cores and compare different
  accretion rates, we perform two sets of simulations which we repeat ten times each
  with varying initial seeds for the perturbations and use the mean of the number of cores.
  The accretion rate is varied by setting the accretion density in the inflow region
  to $n_\text{0} = 25.0$ \ccm, corresponding to $\rho_\text{0} = 9.80\times 10^{-23}$ \gccm,
  and $n_\text{0} = 50.0$ \ccm, corresponding to $\rho_\text{0} = 1.96\times 10^{-22}$ \gccm
  with respective accretion rates of 10.5 and 21.0 $\text{M}_\odot \text{ pc}^{-1} \text{ Myr}^{-1}$.
  For larger accretion rates the number of cores which form are too numerous to be
  reliably distinguished from each other. Therefore, we do not include larger accretion
  rates in our simulations.

\begin{figure}
    \includegraphics[width=1.0\columnwidth]{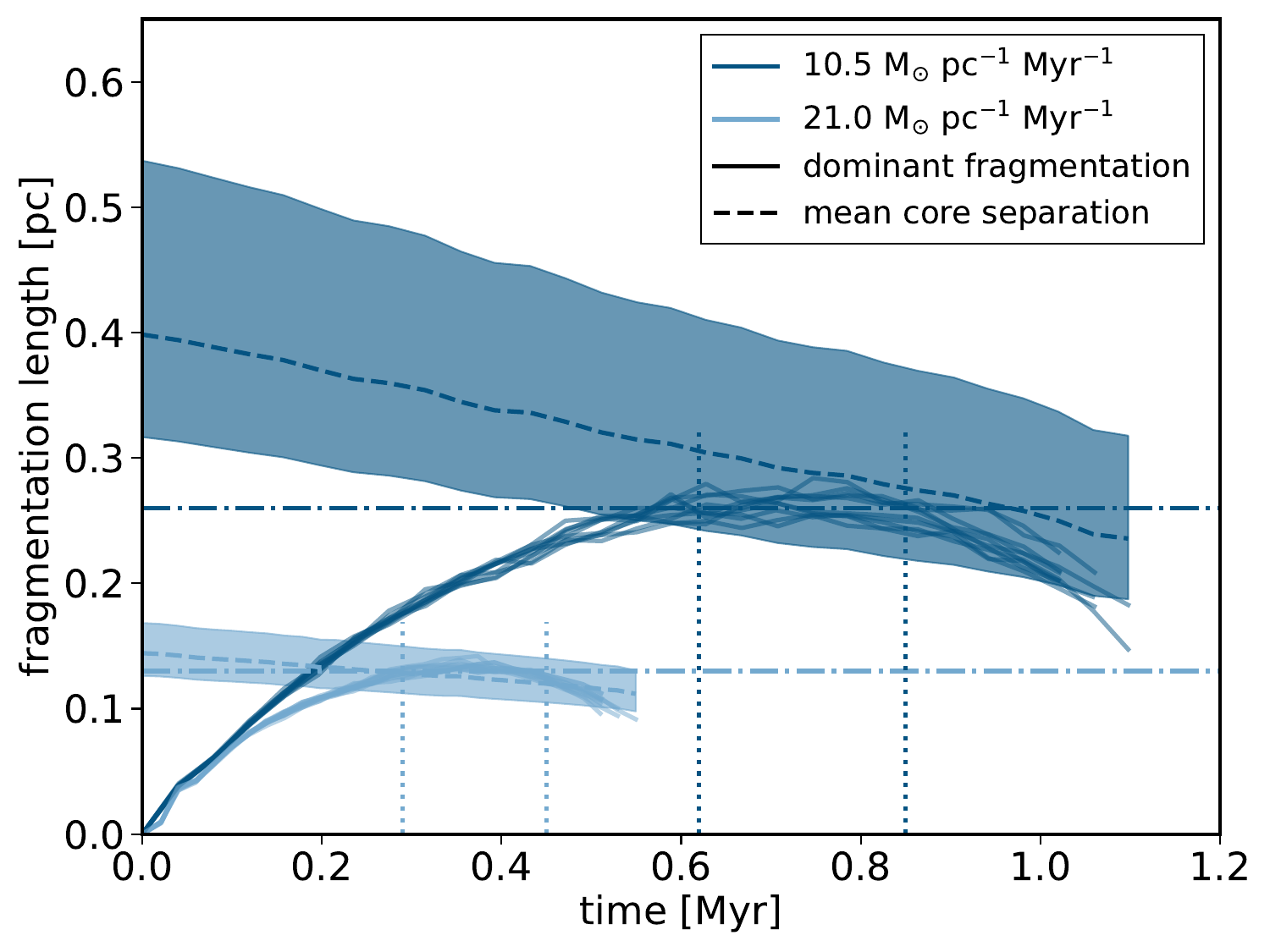}
    \caption{Evolution of the dominant fragmentation length and the core
    separation over time for different accretion rates. The bundle of solid lines
    shows the values of the dominant fragmentation length as calculated from the
    simulations starting with different initial seeds. The dashed lines
    show the mean core separation determined by measuring the mean length of the
    dense inner region over time and dividing it by the mean number of cores.
    The corresponding error bars, given by the filled shaded area, show the
    standard deviation which follows from the variation of the number of cores
    in the different simulations. Both distributions show an overlap close to the
    peak of the dominant fragmentation lengths given by the horizontal dashed-dotted
    lines. We indicate the length of the time period where the dominant
    fragmentation length is close to the peak value by the vertical dotted lines.}
    \label{fig:t_ldom}
  \end{figure}

  The expected dominant fragmentation length is determined for all simulations at every
  output time by calculating the mean central density and mean line-mass of the dense inner
  region and using these values in \autoref{eq:dom}. We show the resulting evolution
  of the fragmentation length in \autoref{fig:t_ldom} as the bundle of semi-transparent
  solid lines. One can see that the dominant fragmentation length rises over time unit it
  reaches a plateau after which it starts to fall again until we stop the simulations due
  to the collapse of a core. The form of the curves follows the evolution of the scale height
  which evolves similar for increasing line-mass and has its maximum at $\fcyl=0.5$. The curves
  all show similar values with some differences due to the initial random seed which grow over
  time. For larger accretion rates, the dominant fragmentation length is shorter and fragmentation
  happens faster which is consistent with analytical models due to the larger central density
  and the smaller scale height.

  We compare the dominant fragmentation length predicted by theory to the mean core
  separation we measure in our simulations. The mean core separation is calculated by
  dividing the length of the dense inner region, the region which is undergoing fragmentation,
  by the number of cores counted at the end of the simulation minus one in order
  to take into account the end cores. We determine the length of the inner dense
  region as described before by locating the outer end of the end cores where the
  column density rises above the mean column density inside the dense inner region.
  Since the inner dense region is not defined at the start of the simulation, we
  use the length of the total inflow region of 1.2 pc as theoretical value of the
  inner region. Note, that this definition also gives us an artificial core separation
  value even before the cores began to form. This however is intended as we do not
  know the point in time when the cores form and as we are interested if we can
  determine it by matching the backtraced core separation to the dominant
  fragmentation length.

  In order to count the number of cores, we determine the number of
  overdensities in the last output file as seen in \autoref{fig:pline}.
  Doing this towards the end of the simulation has the additional advantage of
  more evolved cores which makes them easier to distinguish from small
  fluctuations in the filament. As the cores have varying central
  densities, column densities and line-masses, there is no general good criteria
  for classifying a particular overdensity as a core. Therefore, we use the output
  of the clump finder algorithm employed by \textsc{ramses} to determine the number
  of cores. The algorithm uses a 'watershed' segmentation in order to find connected
  dense regions above a given density threshold. It uses an automatic noise reduction
  and allows for merging of peaks above a given ratio of central to saddle density
  called 'relevance'. More details can be found in \citet{bleuler2015}.
  As we only want to detect the overdensities in the central region, we set the
  density threshold to a typical central density of a filament of $n = 10^5$ \ccm.
  In addition, in order to reduce noise, we set the relevance threshold to 3.0 and
  only use clumps which have reached a minimum mass of 0.1 solar masses. Using this
  method we determine the mean core number over all ten different realisations for
  increasing accretion rate to be 3.0 and 8.3 with a standard deviation of 0.77 and
  1.19, respectively. There is some variation with the number of cores depending
  on how large we set the relevance and the mass threshold with the mass threshold
  showing a larger influence. Adapting a much smaller mass threshold of 0.01 solar
  masses however only increases the number of cores by a factor of around 10\%.

  We plot the mean core separation over time in \autoref{fig:t_ldom} as the
  dashed line together with its standard deviation resulting from the variability
  in core numbers as the shaded area. As the inner dense region collapses along its axis
  the mean core separation decreases over time. Compared to the theoretically expected
  dominant fragmentation length, one can see that for early times the distributions do
  not match, but as soon as the peak of the dominant fragmentation is reached there is
  a direct overlap with the mean core separation after which the dominant fragmentation
  shows a trend of falling below the core separation again. This overlap could indicate
  that cores are generally seeded when the dominant fragmentation reaches its peak.
  However, this is only a first order approach as the dominant fragmentation length
  is constantly changing. Nevertheless, one can see from the broadness of the peak
  of the dominant fragmentation length that the filaments spend a considerable amount of time
  at similar dominant fragmentation lengths indicated by the horizontal dashed-dotted
  lines. For smaller and larger times the change in dominant fragmentation
  length is much faster so any change in line-mass would affect the dominant fragmentation
  length much more. Therefore, perturbations seeded at the peak of the dominant
  fragmentation length could potentially grow close to the seed mode longer than at any
  other point in the filament's evolution. This could lead to an initial core growth which
  is strong enough to exceed other modes at earlier and later times even if the dominant
  fragmentation length is constantly changing.

\begin{figure}
    \includegraphics[width=1.0\columnwidth]{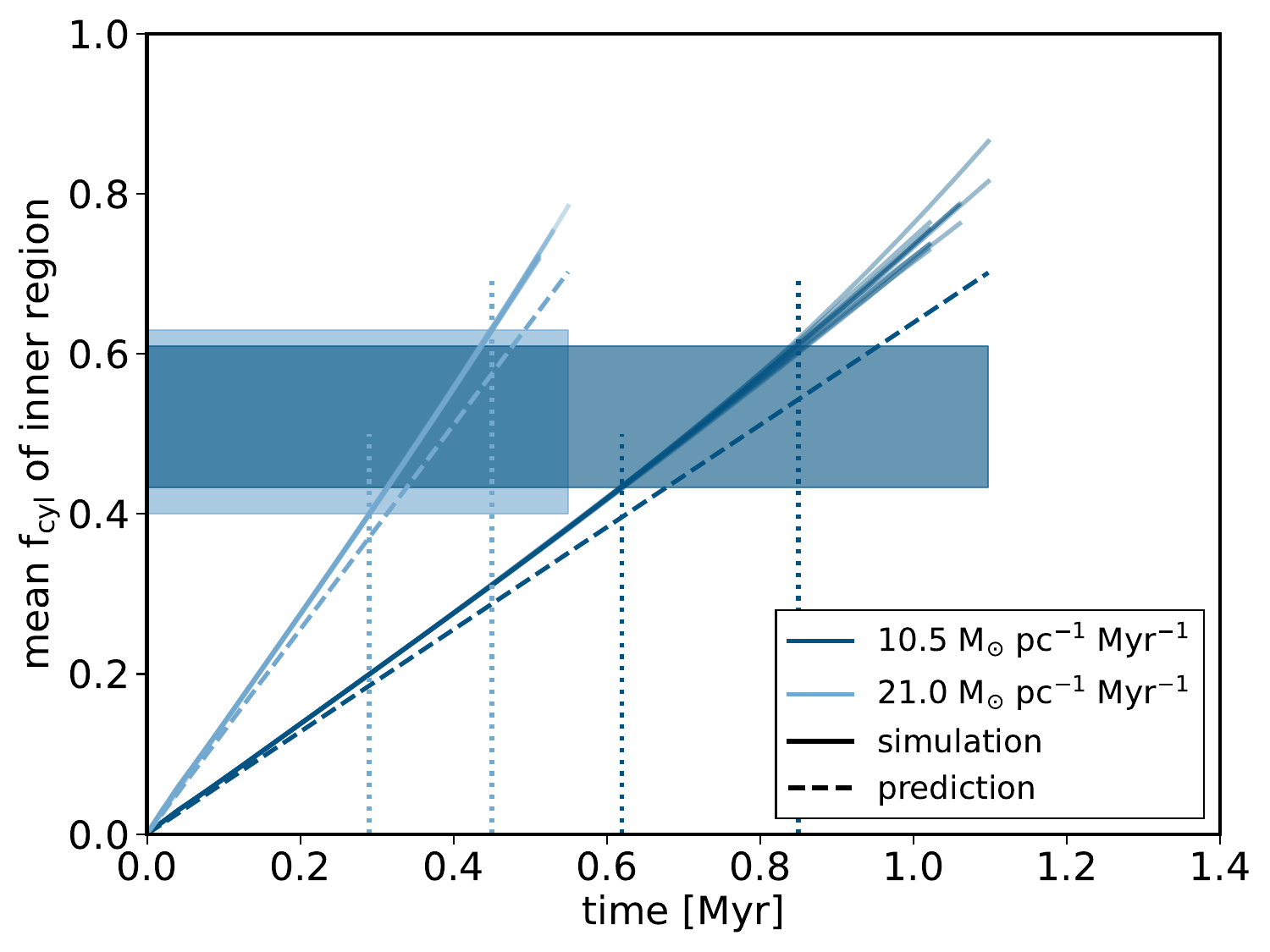}
    \caption{Evolution of the mean line-mass in units of $\fcyl$ of the dense inner
    region over time for different mass accretion rates. The measured mean line-mass
    of all simulations is given by the solid lines and the predicted mean
    line-mass from a constant mass accretion rate as given by the legend is shown by the
    dashed lines. The vertical dotted lines indicate the time period in
    where the dominant fragmentation length is close to the peak value as given
    by \autoref{fig:t_ldom} and the filled shaded areas show the respective
    interpolated values of the mean line-mass.}
    \label{fig:t_fcyl}
  \end{figure}

  In order to verify the compatibility of our theory we measure the time spent
  by the filament at the peak of the curve
  as given by the vertical dotted lines which is from 0.62 Myrs to 0.85 Myrs for the
  low accretion rate and from 0.29 to 0.45 Myrs for the large accretion rate,
  respectively. For both accretion rates we determine the mean e-folding growth time
  The e-folding timescales of the dominant fragmentation length are given by the
  corresponding values in table E.1 of \citet{fischera2012} for the polynomial
  approximation
  \begin{equation}
    \tau_\text{dom}(\fcyl)\cdot\sqrt{4\pi G \rho_c} = \sum_{i=0}^5 a_i \fcyl^{i/3}
    \label{eq:tdom}
  \end{equation}
  and are listed in \autoref{table:constants}. As it depends on the line-mass of
  the filament, we map the respective points in time to the line-mass as can be
  seen in \autoref{fig:t_fcyl} where we plot the evolution of the line-mass over
  time. We show the measured mean line-mass for each simulation as a bundle of
  semi-transparent solid lines together with the expectation if the inner dense
  region would not collapse as the dashes lines. In this case the line-mass would
  increase linearly over time, however, as the inner dense region is shortening
  it increases faster than in the non-collapsing case as more material is
  concentrated in a shorter structure. The vertical dotted lines mark the same
  points in time as in \autoref{fig:t_ldom}. The interpolated mean line-masses at
  the beginning and the end of the time period close to the peak are similar in
  both cases, 0.43 to 0.61 for the low accretion case and 0.40 to 0.63 for the
  large accretion case, and are shown as filled shaded areas. Using the marked
  points in time as bounding cases, we insert the growing mean line-mass and
  corresponding mean central density in \autoref{eq:tdom} and calculate the respective
  mean e-folding timescale over this time period for each simulation separately.
  Finally, we take the mean of these values for the respective accretion rate. For both
  the low and large accretion rate the timescales match well. The time spent close to
  the peak value is given by 0.23 and 0.16 Myrs respectively, while the mean dominant
  e-folding time is 0.17 and 0.11 Myrs. This demonstrates that cores seeded at the peak
  of the curve of the dominant fragmentation length have enough time to grow for more
  than an e-folding time and therefore could exceed earlier and later modes as these
  have less time to grow at their respective dominant fragmentation length. Thus, we
  see a tentative indication that the analytic predictions of the hydrostatic model
  are consistent with the fragmentation in simulated accreting filaments.

  In general, increasing the accretion rate leads to the formation of more
  cores. This is to be expected as a larger accretion rate leads to a larger
  ram pressure onto the filament which increases its central density which
  in turn lowers the dominant fragmentation length. However, from the core
  separation alone one cannot derive the accretion rate as the same ram
  pressure can be achieved by varying numbers of accretion density and
  velocity. Nevertheless, one can use the observed core separation as an
  indicator of the ambient pressure. As one can see in \autoref{fig:t_ldom},
  at least for larger accretion rates, the line-mass of the filament grows
  faster than the central dense region contracts which results in a
  flattening of the measured core separation for larger values of the
  filament line-mass. If one now assumes that the cores are seeded at
  values of $\fcyl \approx 0.4-0.6$, one can in principle estimate the
  ambient pressure of the surrounding medium leading to a corresponding
  dominant fragmentation length.

\section{Discussion and conclusions}
\label{sec:discussion}

  While our simulations show that a continuous accretion does prevent the
  formation of dominant end cores, several additional questions remain.

  An intriguing puzzle is the tendency of observations to find narrower
  core separations in large line-mass filaments than what is predicted by
  theory \citep{andre2010, kainulainen2013, teixeira2016}. Our analysis of
  expected fragmentation lengths presented in \autoref{fig:t_ldom} shows that
  for large times the dominant fragmentation decreases faster than
  the inner region contracts. Therefore, in contrast to observations, the
  predicted fragmentation length is usually shorter than what we measure in
  simulations. This could be an indication that such a hydrostatic model is
  not applicable for all filaments.

  However, the shapes of the curves formed by the dominant fragmentation length
  shown in \autoref{fig:t_ldom} are very likely to change depending on the mass
  accretion onto the filament and turbulence within. We presented a very idealised
  model with a constant mass accretion rate without initial turbulent motions.
  A more realistic case with varying mass accretion rate and turbulent motions
  in the accreted material is very likely to shift the peak of the curves.
  Thus, the values of $\fcyl \approx 0.4-0.6$ at which the cores are likely
  seeded in our simulations should not be applied necessarily to all
  accreting filaments. Nevertheless, if the dominant fragmentation length shows
  a peak, due to the timescale argument presented in \autoref{sec:cores}
  it is likely that cores will form with a preferred separation close
  to the peak value.

  A surprising side effect of the model is the creation of outflows at
  the filament ends. This effect however could be reduced by non-homogeneous
  inflows with low density at the shock front ends. Due to
  their low density and mass loading, these outflow would be very hard to
  detect and as far as we know there is no observation of such an phenomena
  to date. A possible tracer of the outflow would be the shock heating or
  chemical shock traces in the ambient medium as the material streams out
  with large velocities.

  Finally, it will be interesting to explore whether our results remain valid
  in the case of planar accretion flows instead of radial geometries.
  Depending on the extent, a planar shock will not create a filamentary
  morphology but lead to a sheet between the shock fronts. However, as soon
  as enough material has been concentrated and gravity takes over, this
  sheet could collapse and be collected into a filamentary structure. We
  will explore if this mode of filament formation shows the same effect on
  end core formation as presented here in a future study.

  To summarise, we have presented a numerical study on the simultaneous
  filament and end core formation. We have shown that while end cores
  do form at the ends of the central dense region, they do not dominate over
  the cores formed by regular perturbation growth. This is due to the material
  which is replenished continuously from the inflow outside of the contracting
  inner region which leads to a reduced gravitational acceleration at its ends
  and a soft edge morphology. Comparing the measured core separation
  in our idealised simulations to the dominant fragmentation length shows a
  tentative agreement that cores are seeded at $\fcyl \approx 0.4-0.6$. The
  comparison is not straight-forward due to a continuously shifting dominant
  fragmentation length over time, however an accreting filament spends
  a considerable amount of time in this line-mass region where the dominant
  fragmentation length only changes slightly. This means that cores seeded
  at these values have about an e-folding timescale to grow and therefore
  exceed later core formation modes. Independent of the core separation,
  we presented a valid mechanism to form a filament morphology which is not
  dominated by the edge effect as was the goal of the paper.

\section*{Acknowledgements}

  We thank the referee for greatly improving the clarity and readability
  of the paper. We thank the whole CAST group for helpful comments and
  discussions. This research was supported by the Excellence Cluster ORIGINS
  which is funded by the Deutsche Forschungsgemeinschaft (DFG, German Research
  Foundation) under Germany´s Excellence Strategy – EXC-2094 – 390783311.

\section*{Data Availability}

The data underlying this article will be shared on reasonable request to the corresponding author.



\bibliographystyle{mnras}
\bibliography{Edge.bib}





\bsp	
\label{lastpage}
\end{document}